\documentclass[10pt,twocolumn,showpacs]{revtex4}
\usepackage[T1]{fontenc}
\usepackage[latin9]{inputenc}
\usepackage{color}
\usepackage{textcomp}
\usepackage{amsmath}
\usepackage{amssymb}
\usepackage{graphicx}
\usepackage{esint}
\usepackage{dcolumn}
\usepackage{bm}
\usepackage{wrapfig}
\usepackage{subfigure}
\usepackage{ucs}
\usepackage{lineno}
\usepackage{lmodern}
\usepackage{epstopdf}
\usepackage{epsfig,psfrag}
\DeclareGraphicsExtensions{.pdf,.eps,.png,.jpg,.mps}



\begin{document}

\title{Zero temperature non-plateau magnetization and magnetocaloric effect in an Ising-XYZ diamond chain structure }

\author{J. Torrico$^{1}$, M. Rojas$^{2}$, S. M. de Souza$^{2}$ and Onofre Rojas$^{2}$ }

\affiliation{$^{1}$Instituto de Física, Universidade Federal de Alagoas, 57072-970, Maceió, AL, Brazil}

\affiliation{$^{2}$Departamento de Física, Universidade Federal de Lavras, 37200-000, Lavras, MG, Brazil}

\begin{abstract}
Zero temperature non-plateau magnetization is a peculiar property of a quantum spin chain and it sometimes appears due to different gyromagnetic factors. In this study, we illustrate a quite unusual nonplateau magnetization property driven by XY-anisotropy in an Ising-XYZ diamond chain. Two particles
with spin-1/2 are bonded by XYZ coupling and they are responsible for the emergence of non-plateau magnetization. These two quantum operator spins are bonded to two nodal Ising spins and this process is repeated infinitely to yield a diamond chain structure. Due to the non-plateau magnetization property, we focus our discussion on the magnetocaloric effect of this model by presenting the isentropic curves and the Grüneisen parameters, as well as showing the regions where the model exhibits an efficient magnetocaloric effect. Due to the existence of two phases located very close to each other, the strong XY-anisotropy exhibits a particular behavior with a magnetocaloric effect, with a wider interval in the magnetic field, where the magnetocaloric effect is efficient.
\end{abstract}
\maketitle

\section{Introduction}

The magnetocaloric effect (MCE) occurs when a conventional magnetic material is heated, so the magnetic field is then turned on, before cooling down and the magnetic field shuts  down, where this cyclic property can be used for lowering the temperature of a system in a similar manner to a  conventional refrigerator powered by a vapor cycle. This property is important because of its  potential applications in domestic and industrial refrigeration.
 Several materials have discovered in the past decade that exhibit this property with MCE at room temperature, such as $\mathrm{Mn}_{4}\mathrm{FeGe}_{3-x}\mathrm{Si}_{x}$ (see \cite{Halder} and the reference therein). The MCE above room temperature also occurs in
 $\mathrm{La}_{0.67}\mathrm{Sr}_{0.33}\mathrm{MnO}_{3}$ manganite~\cite{rostamnejadi}. Other interesting materials with a very strong MCE property have been tuned by using  $\mathrm{Mn}_{1-x}\mathrm{Fe}_{x}\mathrm{As}$ at ambient pressure~\cite{Nat-campos}. Theoretically, the entropy
and the cooling rate of the antiferromagnetic spin-$1/2$ XXZ chain have been  investigated under an adiabatic demagnetization process using the
quantum transfer-matrix technique~\cite{Trippe}. The exact results for  the Heisenberg chains were used to confirm the numerically
 exact diagonalization as well as quantum Monte Carlo simulations, which showed that the model exhibited a large MCE close to field-induced quantum
phase transitions~\cite{Trippe}.

Recently,  materials have been discovered  such as $\mathrm{Cu}_{3}(\mathrm{CO}_{3})_{2}(\mathrm{OH})_{2}$ \cite{kikuchi03}, known as azurite, for which an interesting quantum antiferromagnetic model can be  described by a  Heisenberg model based on a generalized diamond chain. According to  experimental
measurements \cite{rule,kikuchi}, the $1/3$ magnetization plateau and a  double peak were observed in both  the magnetic susceptibility and specific heat. It should be noted that the dimer (interstitial sites) exchange is much stronger than those at nodal sites in the XY axes. The  $z$ component of the dimer interaction is much higher than the rest, so  it can be represented well as an exactly solvable Ising--Heisenberg model. Several interesting theoretical
investigations have been reported of  diamond chain models, such as the approximated Ising--Heisenberg model \cite{canova06,ananikian,chakh}, where the experimental data based  on the magnetization plateau agreed with the  theoretical results, as well as that studied by Honecker et al.~\cite{honecker}. Theoretical investigations of the magnetization property have also been  considered for mixed spins based on an Ising--Heisenberg diamond chain \cite{mlis,abga}. Furthermore, the thermodynamics of the Ising--Heisenberg model based on the diamond-like chain have been discussed widely (\cite{canova06,lisnii-11,vadim,valverde,dyan}). The  magnetic properties of an Ising--Heisenberg diamond chain have also been considered  with four spin interaction \cite{lgali}, spin-1 plateau \cite{ana}, and  mixed spin with a biquadratic spin interaction~\cite{deso}.

Some compounds such as the single-chain magnet, $[{(\mathrm{CuL})_{2}\mathrm{Dy}}{\mathrm{Mo}(\mathrm{CN})_{8}}]\cdot2\mathrm{CH}_{3}\mathrm{CN}\cdot \mathrm{H}_{2}\mathrm{O}$ \cite{visinescu,van-prb10} have an interesting spin model with different Land\'{e} g-factors, which can be described well by an Ising--Heisenberg spin chain, thereby leading to  non-plateau curve magnetization at zero temperature. Theoretical investigations of  non-plateau magnetization with different Land\'{e} g-factors have been performed based on the exact analysis approach to show this effect \cite{bellucci,ohany},
which is almost imperceptible in  the magnetization curve due to a very small difference between the  Land\'{e} g-factors for the magnetic ions.
Similar observations have been made of  the enhanced MCE for an exactly solvable spin-$1/2$ Ising--Heisenberg diamond chain with different Land\'{e} g-factors~\cite{gali} and its variant \cite{gali,galisova-2014,canova09}.

The remainder of this  paper is organized as follows. In Section~\ref{sec2},  we present the Ising-XYZ model based on a diamond chain and we discuss the non-plateau magnetization at zero temperature. In Section~\ref{sec3}, we discuss the entropy and MCE property  using the exact thermodynamic
solution of the model. Finally, we give our concluding remarks in Section~\ref{sec4}.\vspace{0.5mm}

\section{The Ising-XYZ chain}\label{sec2}

The motivation to study the Ising-XYZ diamond chain model\cite{torrico} is based on some recent works, such as the experiments of the natural mineral azurite\cite{kikuchi03}.

\begin{figure}
\includegraphics[scale=0.5]{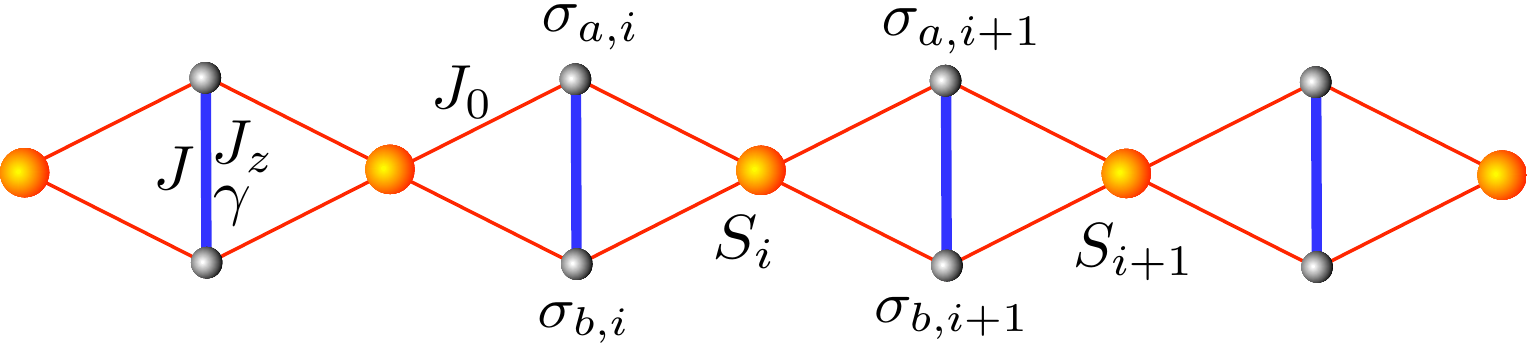}\caption{\label{fig:Schem}(Color Online) Schematic representation of Ising-XYZ
chain on diamond structure. By $S_{i}$ spins corresponds to Ising
spins, and $\sigma_{a(b),i}$ denotes the Heisenberg spins.}
\end{figure}

Now let us assume the Ising-XYZ diamond chain structure as schematically illustrated in figure \ref{fig:Schem}. Where $S_{i}$ represents the Ising spin-1/2, and $\sigma_{a(b),i}^{\alpha}$ denoting the Heisenberg spin-1/2, assuming $\alpha=\{x,y,z\}$.

Thus, the Ising-XYZ Hamiltonian can be expressed as
\begin{alignat}{1}
H= & -\sum_{i=1}^{N}\left[J(1+\gamma)\sigma_{a,i}^{x}\sigma_{b,i}^{x}+J(1-\gamma)\sigma_{a,i}^{y}\sigma_{b,i}^{y}+\right.\nonumber \\
 & +J_{z}\sigma_{a,i}^{z}\sigma_{b,i}^{z}+J_{0}(\sigma_{a,i}^{z}+\sigma_{b,i}^{z})(S_{i}+S_{i+1})+\nonumber \\
 & \left.+h(\sigma_{a,i}^{z}+\sigma_{b,i}^{z})+\frac{h}{2}(S_{i}+S_{i+1})\right],\label{eq:Hamt}
\end{alignat}
where $J$ corresponds to $xy$-axes exchanges and with $\gamma$ being the XY-anisotropy, $J_{z}$ corresponds to Heisenberg spins exchange, whereas $J_{0}$ denotes Ising-Heisenberg spins exchange, and $h$ corresponds to the external magnetic field along the $z$-axis.

After performing the diagonalization in the XYZ dimer, of the Hamiltonian \eqref{eq:Hamt} for a given elementary cell, we obtain the following eigenvalues, assuming $\mu=S_{i}+S_{i+1}$,
\begin{alignat}{1}
\varepsilon_{1,4}= & -h\frac{\mu}{2}-\frac{J_{z}}{4}\pm\Delta(\mu),\\
\varepsilon_{2,3}= & -h\frac{\mu}{2}\mp\frac{J}{2}+\frac{J_{z}}{4},
\end{alignat}
wherein $\Delta(\mu)=\sqrt{\left(h+J_{0}\mu\right)^{2}+\tfrac{1}{4}J^{2}\gamma^{2}}.$

With the corresponding eigenvectors in standard basis $\{|\begin{smallmatrix}-\\
-
\end{smallmatrix}\rangle,|\begin{smallmatrix}-\\
+
\end{smallmatrix}\rangle,|\begin{smallmatrix}+\\
-
\end{smallmatrix}\rangle,|\begin{smallmatrix}+\\
+
\end{smallmatrix}\rangle\}$ are given respectively by
\begin{alignat}{1}
|\varphi_{1}\rangle & =c_{+}\left(\alpha_{+}|\begin{smallmatrix}+\\
+
\end{smallmatrix}\rangle+|\begin{smallmatrix}-\\
-
\end{smallmatrix}\rangle\right),\label{eq:n-Mg1}\\
|\varphi_{2}\rangle & =\tfrac{1}{\sqrt{2}}\left(|\begin{smallmatrix}-\\
+
\end{smallmatrix}\rangle+|\begin{smallmatrix}+\\
-
\end{smallmatrix}\rangle\right),\label{eq:0-mg1}\\
|\varphi_{3}\rangle & =\tfrac{1}{\sqrt{2}}\left(|\begin{smallmatrix}-\\
+
\end{smallmatrix}\rangle-|\begin{smallmatrix}+\\
-
\end{smallmatrix}\rangle\right),\label{eq:0-mg2}\\
|\varphi_{4}\rangle & =c_{-}\left(\alpha_{-}|\begin{smallmatrix}+\\
+
\end{smallmatrix}\rangle+|\begin{smallmatrix}-\\
-
\end{smallmatrix}\rangle\right),\label{eq:n-Mg2}
\end{alignat}
where $\alpha_{\pm}=\frac{-J\gamma}{2h+2J_{0}\mu\pm2\Delta(\mu)}$, and $c_{\pm}=\frac{1}{\sqrt{1+\alpha_{\pm}^{2}}}$.

These eigenstates play an important role in discussion of the geometrically frustrated regions and non-plateau magnetization.

\subsection{Phase diagram of Ising-XYZ on diamond chain}

Here, we present a brief review, concerning the phase diagram at zero temperature, which was previously studied in reference \cite{torrico}. The phase diagram in figure 2b and 3a of reference \cite{torrico}, were illustrated three phases.

Two explicit representation of Ising spin ferromagnetic with Heisenberg spin modulated ferromagnetically or simply denoted by modulated ferromagnetic ($FMF$) phase, are expressed below
\begin{alignat}{1}
|FMF_{1}\rangle= & \overset{N}{\underset{i=1}{\prod}}|\varphi_{4}\rangle_{i}\otimes|+\rangle_{i},\label{eq:FMF1}\\
|FMF_{2}\rangle= & \overset{N}{\underset{i=1}{\prod}}|\varphi_{4}\rangle_{i}\otimes|-\rangle_{i},\label{eq:FMF2}
\end{alignat}
whereas the corresponding ground-state energies per unit cell are given by
\begin{alignat}{1}
E_{FMF_{1}}= & -\frac{h}{2}-\frac{J_z}{4}-\sqrt{\left(h+J_{0}\right)^{2}+\tfrac{1}{4}J^{2}\gamma^{2}},\label{eq:EFM1}\\
E_{FMF_{2}}= & \frac{h}{2}-\frac{J_z}{4}-\sqrt{\left(h-J_{0}\right)^{2}+\tfrac{1}{4}J^{2}\gamma^{2}}.\label{eq:EFM2}
\end{alignat}

The total magnetization per unit cell at zero temperature, is defined by $M=-\tfrac{\partial E_{0}}{\partial h}$ with $E_{0}$ being the ground state energy per unit cell of the system. Thus, the normalized magnetization for each phases become

\begin{alignat}{1}
\tfrac{M_{FMF_{1}}}{M_{s}}=-\tfrac{\partial E_{FMF_{1}}}{M_{s}\partial h}= & \tfrac{1}{3}+\tfrac{2}{3}\tfrac{\left(h+J_{0}\right)}{\sqrt{\left(h+J_{0}\right)^{2}+\frac{\gamma^{2}J^{2}}{4}}},\label{eq:Mg1}\\
\tfrac{M_{FMF_{2}}}{M_{s}}=-\tfrac{\partial E_{FMF_{2}}}{M_{s}\partial h}= & -\tfrac{1}{3}+\tfrac{2}{3}\tfrac{\left(h-J_{0}\right)}{\sqrt{\left(h-J_{0}\right)^{2}+\frac{\gamma^{2}J^{2}}{4}}},\label{eq:Mg4}
\end{alignat}
where $M_{s}$ means total saturated magnetization.

Note that the eigenstates \eqref{eq:n-Mg1} and \eqref{eq:n-Mg2} depends of $\gamma$, and for $\gamma\ne0$ the magnetization plateau vanishes. It is worth to mention that $FMF_{1}$ and $FMF_{2}$ are degenerated state at zero magnetic field.

One explicit representation of Ising spin ferromagnetic with Heisenberg spin antiferromagnetic ($FAF$) phase, which is expressed below

\begin{alignat}{1}
|FAF\rangle= & \overset{N}{\underset{i=1}{\prod}}|\varphi_{2}\rangle_{i}\otimes|+\rangle_{i},
\end{alignat}
while its ground state energy is given by
\begin{alignat}{1}
E_{FAF}= & -\frac{J+h}{2}+\frac{J_z}{4}.\label{eq:FAF}
\end{alignat}

Whereas the magnetization of FAF state is given simply by $M_{FAF}/M_{s}=1/3$.

\subsection{Non-plateau magnetization}

Usually, the magnetization of Ising and Heisenberg models exhibit plateaux at zero temperature, although one can observe a non-plateau behavior due to the different gyromagnetic Landé factors\cite{Vadim-prb15}. However, here we display a different non-plateau behavior at zero temperature, which is highly influenced by XY-anisotropy parameter $\gamma$.

A simple way how to obtain the magnetization at zero temperature can be obtained from the ground state energy given by eqs. \eqref{eq:EFM1}, \eqref{eq:EFM2} and \eqref{eq:FAF}.

\begin{figure}
\includegraphics[scale=0.2]{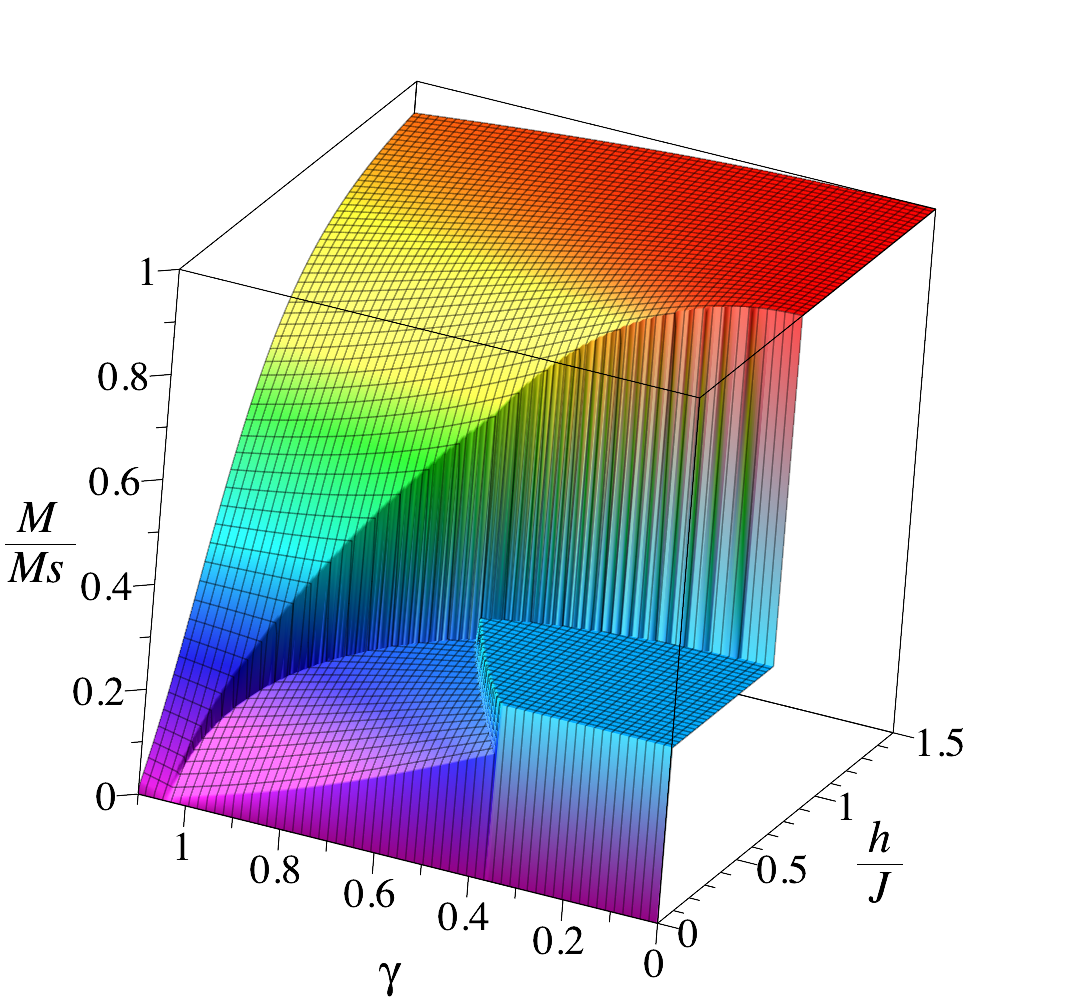}\caption{\label{fig:Mag-zero}(Color Online) Total magnetization $M/M_{s}$ versus external magnetic field $h/J$ and XY-anisotropy parameter $\gamma$ at zero temperature for fixed model parameters $J_{0}/J=-0.3$ and $J_{z}/J=0.3$.}
\end{figure}

In Fig. \ref{fig:Mag-zero}, we illustrate the total magnetization $M/M_{s}$ versus external magnetic field $h/J$ and the XY-anisotropy parameter $\gamma$ at zero temperature assuming fixed parameters $J_{0}/J=-0.3$ and $J_{z}/J=0.3$. A perfect plateau region corresponds to $FAF$ region with 1/3 magnetization plateau, while the non-plateau regions correspond to the $FMF$ region. The lower non-plateau magnetization corresponds to the $FMF_{2}$ region and the larger non-plateau region corresponds to $FMF_{1}$, with $M/M_{S}$ given by eqs.\eqref{eq:Mg1} and the curvature is a consequence of XY-anisotropy parameter $\gamma$. The magnetization curve for $\gamma=0$ drops in a perfect plateau at $M/M_{s}=1/3$. Whereas the larger non-plateau curve represents $FMF_{1}$ given by eq.\eqref{eq:Mg4}, note that for $\gamma=0$, this curve becomes a saturated plateau $M/M_{s}=1$.For $\gamma\gtrsim1$, this plateau vanishes definitely according to eqs.( \ref{eq:Mg1}-\ref{eq:Mg4}).

\begin{figure}
\includegraphics[scale=0.3]{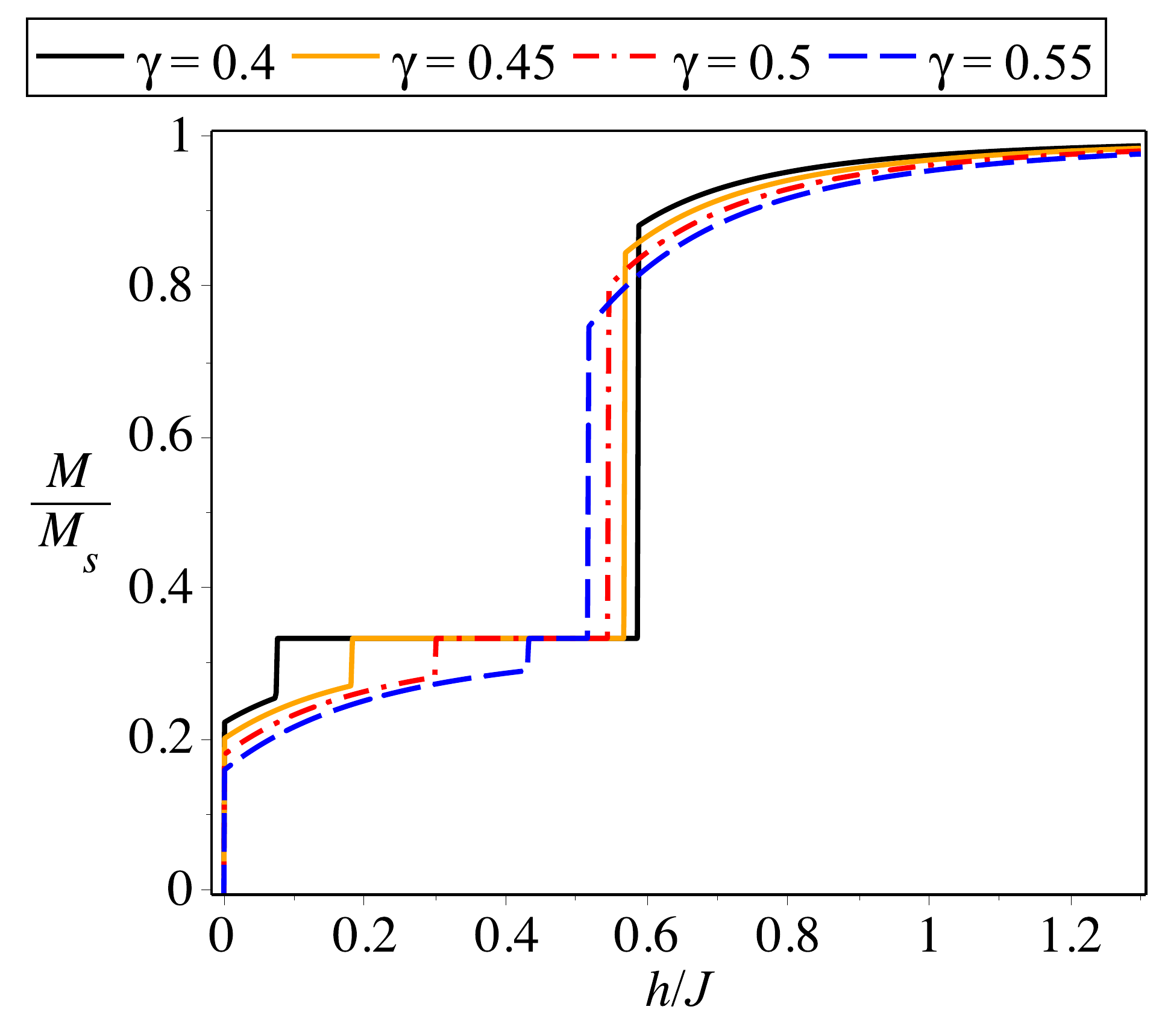}\caption{\label{fig:Mag-XYZ}(Color Online) Total magnetization versus external magnetic field \textcolor{black}{$h/J$ at zero temperature for fixed parameters }$J_{0}/J=-0.3$ and $J_{z}/J=0.3$ and various values of XY-anisotropy $\gamma$.}
\end{figure}

In figure \ref{fig:Mag-XYZ}, we also display a typical magnetization $M/M_{s}$ as a function of $h/J$, considering fixed value $J_{0}/J=-0.3$ and $J_{z}/J=0.3$. As soon as the XY-anisotropy increases, the 1/3 magnetization plateau shrinks ($FAF$ region). Once again, we display two non-plateau magnetization due to the influence of $\gamma$ parameter, when magnetic field is increased ($h/J$), the largest curve leads asymptotically to a saturated magnetization, which corresponds to the $FMF_{1}$ region, while the lower non-plateau curve corresponds to $FMF_{2}$. The $1/3$ plateau dissapears when
$\gamma\rightarrow0.6$ , this one shows the non-plateau curve is strongly dependent of $\gamma$ parameter.

\begin{figure}
\includegraphics[scale=0.38]{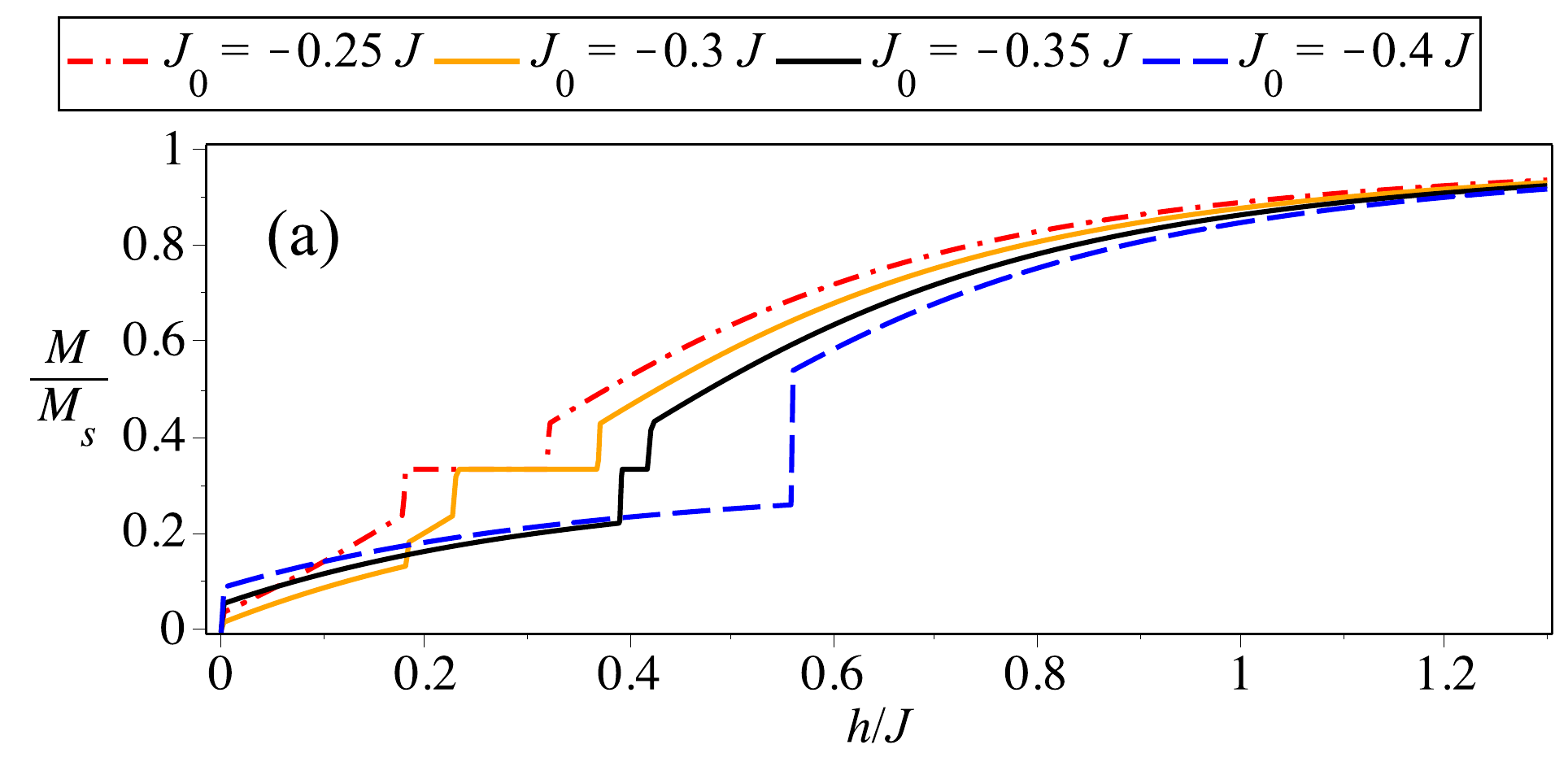} \includegraphics[scale=0.38]{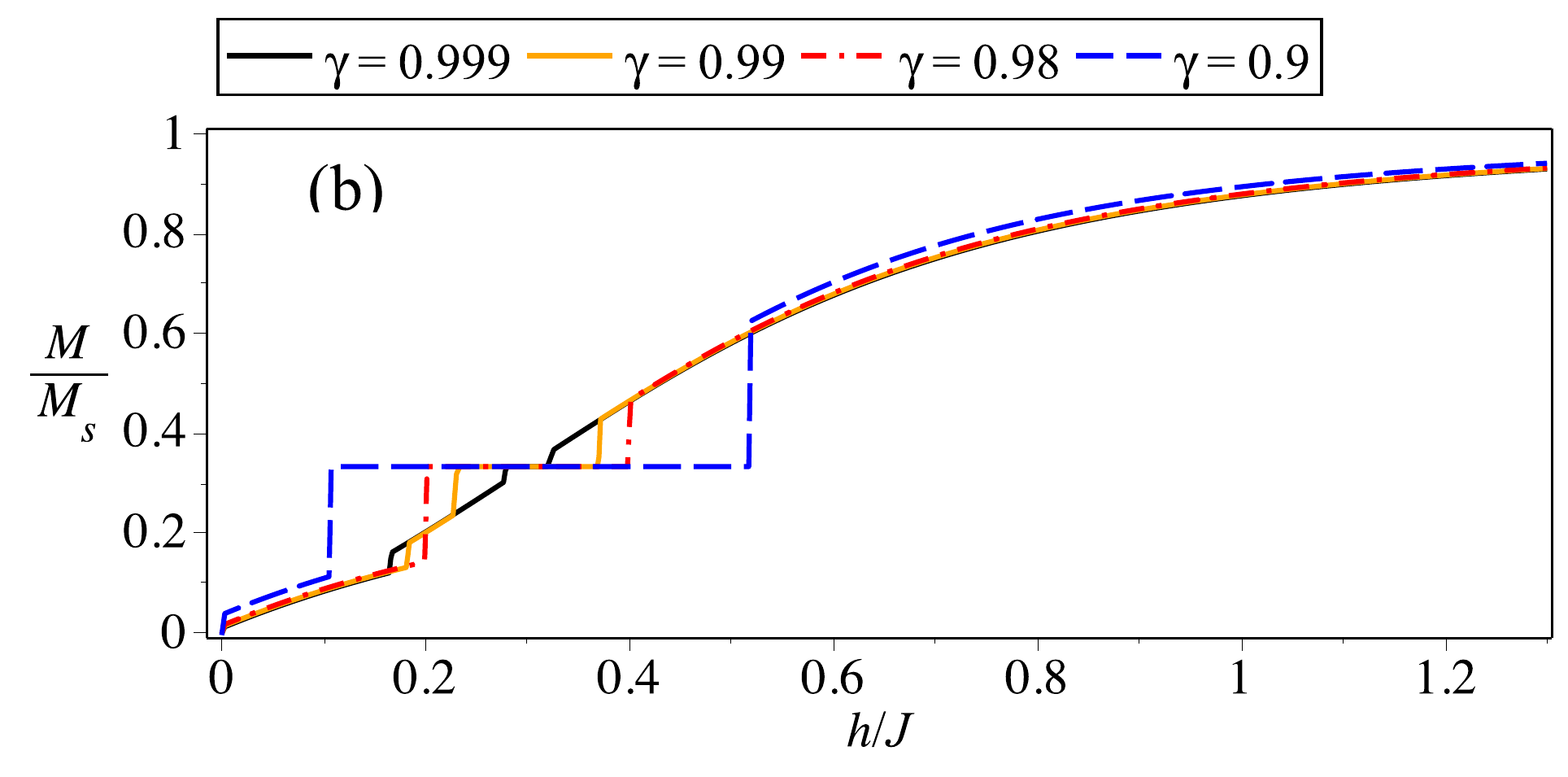}\caption{\label{fig:Mag-plat}(Color Online) Total magnetization versus external
magnetic field \textcolor{black}{$h/J$ at zero temperature}, assuming $J_{z}/J=0$. (a) For different values of $J_{0}/J$ and $\gamma=0.99$. (b) For different values of $\gamma$ and $J_{0}/J=-0.3$. }
\end{figure}

For the limiting case of Ising-XY diamond chain ($J_{z}/J=0$), is also illustrated a typical magnetization plateau in figure \ref{fig:Mag-plat}(a) as a function of magnetic field $h/J$ for several values of $J_{0}/J$ and fixed $\gamma=0.99$, where is illustrated a perfect plateau only at the 1/3 magnetization and this plateau vanishes for $J_{0}/J$ lower than $-0.35$. A similar plot is displayed in figure \ref{fig:Mag-plat}(b) as a function of magnetic field $h/J$, for several values of $\gamma$ and fixed $J_{0}/J=-0.3$. Where we can observe for $\gamma\rightarrow1$, the 1/3 magnetization plateau disappears definitely. In general, for $\gamma\neq0$ the saturation magnetization occurs only in the asymptotic limit of high magnetic field.

\section{Thermodynamics}

The free energy of Ising-XYZ diamond chain was obtained exactly in reference \cite{torrico}, using the decoration transformation\cite{Syozi,Fisher,phys-A-09,strecka pla} together with the usual transfer matrix technique. Thus, the partition function is given by $\mathbb{\mathcal{Z}}={\rm tr}\left({\rm e}^{-\beta H}\right),$ where $\beta=1/k_{B}T$, with $k_{B}$ being the Boltzmann constant and $T$ is the absolute temperature, whereas the Hamiltonian $H$ is given by eq.\eqref{eq:Hamt}. Using the transfer matrix approach, the eigenvalues of the model has been obtained in reference\cite{torrico} by,
\begin{equation}
\lambda_{\pm}=w(1)+w(-1)\pm\sqrt{\left(w(1)-w(-1)\right)^{2}+4w(0)^{2}}.
\end{equation}
where the Boltzmann factor is expressed by
\begin{equation}
w(\mu)=2{\rm e}^{\frac{\beta\mu h}{2}}\left[{\rm e}^{-\frac{\beta J_{z}}{4}}\cosh\left(\tfrac{\beta J}{2}\right)+{\rm e}^{\frac{\beta J_{z}}{4}}\cosh\left(\beta\Delta(\mu)\right)\right].
\end{equation}

Thus, the free energy per unit cell in thermodynamic limit becomes
\begin{equation}
f=-\frac{1}{\beta}\underset{N\rightarrow+\infty}{\lim}\frac{\ln\mathcal{Z}}{N}=-\frac{1}{\beta}\ln\lambda_{+}.\label{eq:free-enr}
\end{equation}

Thus, we are able to calculate the entropy and the magnetocaloric effect of the model.

\subsection{Entropy}

In what follows, we will analyze the behavior of entropy for the same set of parameters as in ground-state phase diagram shown in figure 2b and 3a of reference \cite{torrico}. The entropy can be easily obtained from the free energy using the relation $\mathcal{S}=-\frac{\partial f}{\partial T}$.

\begin{figure}
\includegraphics[scale=0.16]{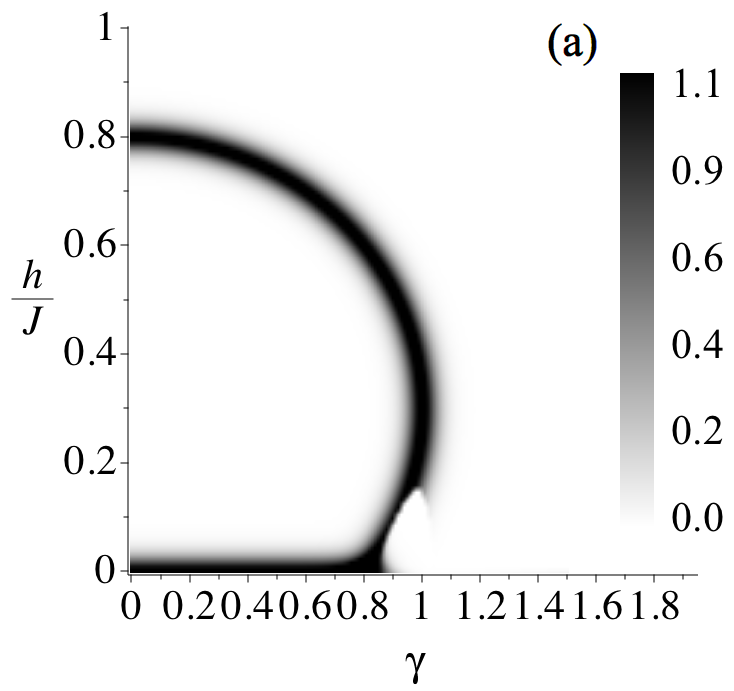}\hspace{2mm}\includegraphics[scale=0.16]{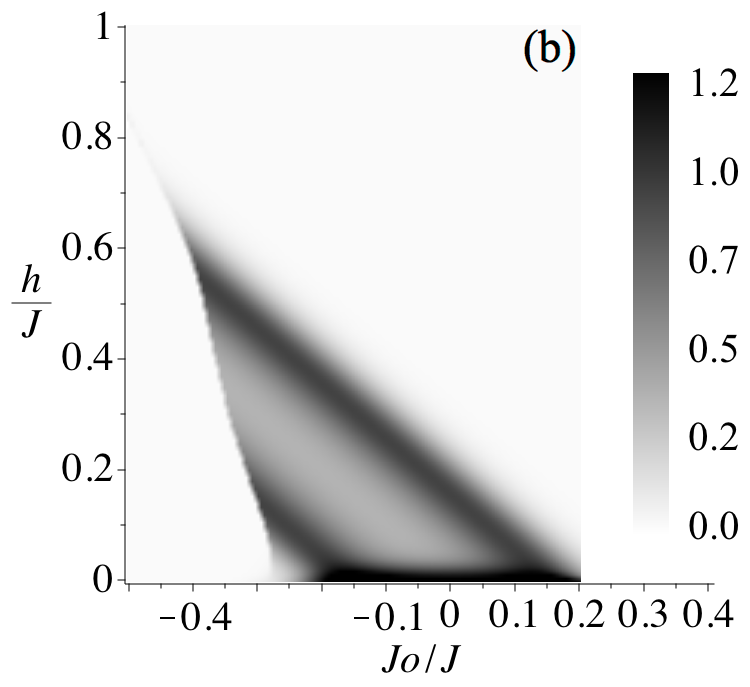}\caption{\label{fig:Denspl-Entro}Density-plot of the entropy $\mathcal{S}$ for the low temperature limit $k_{B}T/J=0.01$. (a) In the plane $\gamma$-$h/J$ by assuming $J_{0}/J=-0.3$ and $J_{z}/J=0$. (b) In the plane of $J_{0}/J$-$h/J$ by assuming $\gamma=0.95$ and $J_{z}/J=0$.}
\end{figure}

In figure \ref{fig:Denspl-Entro}(a), we illustrate the density plot of entropy in terms of $\gamma$ and $h/J$, considering fixed $J_{z}/J=0$ and $J_{0}/J=-0.3$, the plot was performed in the low temperature limit $T/J=0.01$. The dark region corresponds to higher entropy while the white region corresponds to zero entropy. Furthermore, one can observe a similar density-plot of the entropy in figure \ref{fig:Denspl-Entro}(b), in terms of $J_{0}/J$ and $h/J$, once again, in the low temperature limit $T/J=0.01$ for a fixed value of $\gamma=0.95$ and $J_{z}/J=0$. In this panel, we observe the entropy in $FMF_{2}$ region leads abruptly to zero entropy, while, the boundary between $FAF$ and $FMF_{1}$ region leads smoothly to zero entropy. Besides, we observe also that for zero magnetic field, the darkest region is related to the residual entropy $\mathcal{S}=k_{B}\ln(2)$, where the entropy increases abruptly,
as soon as the temperature increases in the low temperature region.

\subsection{Magnetocaloric effect}

Recently, it has been demonstrated that several frustrated spin systems may exhibit an enhanced magnetocaloric effect (MCE) during the adiabatic demagnetization process, which could be of great importance to study the low-temperature magnetic refrigeration. Due to this fact, let us also investigate the adiabatic demagnetization process of the Ising-XYZ diamond chain under the adiabatic (isentropic) conditions. Here, we are interested in studying the isentropes (entropy levels) in the plane ($h/J$ - $k_{B}T/J$) and Grüneisen parameter $\Gamma_{h}$,

\begin{figure}
\includegraphics[scale=0.1]{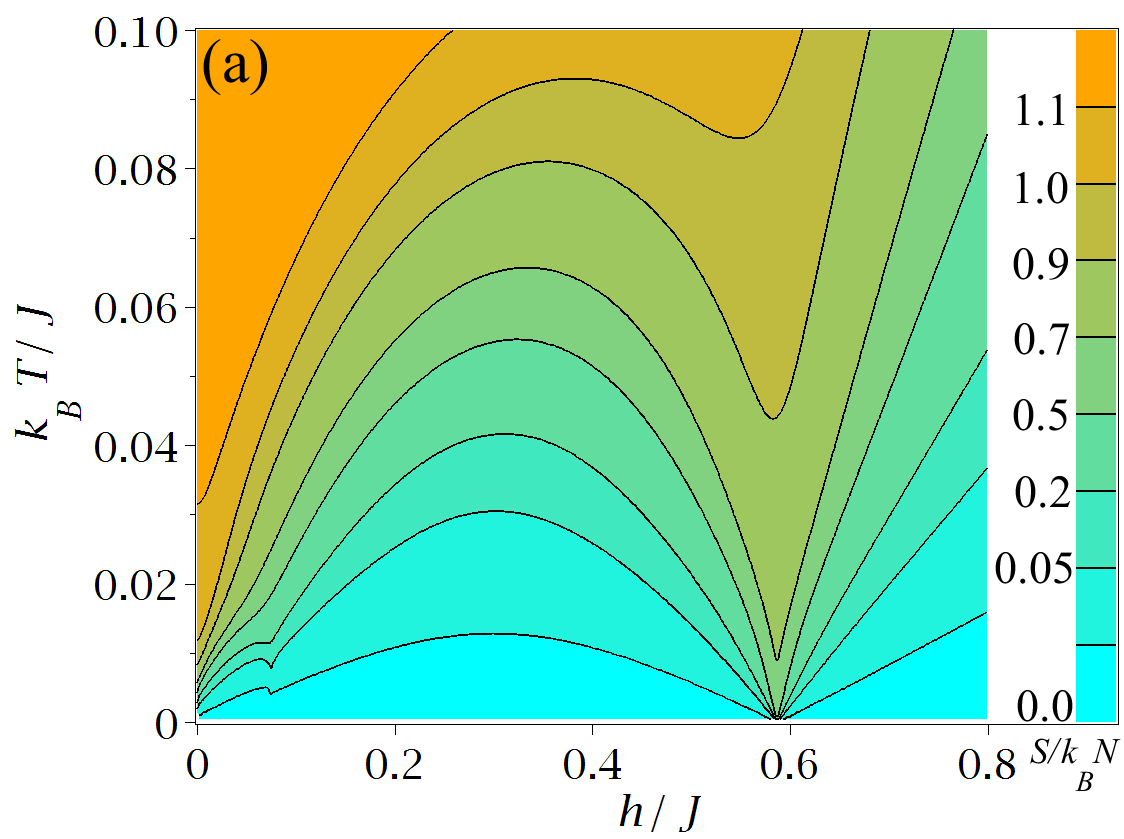}\includegraphics[scale=0.1]{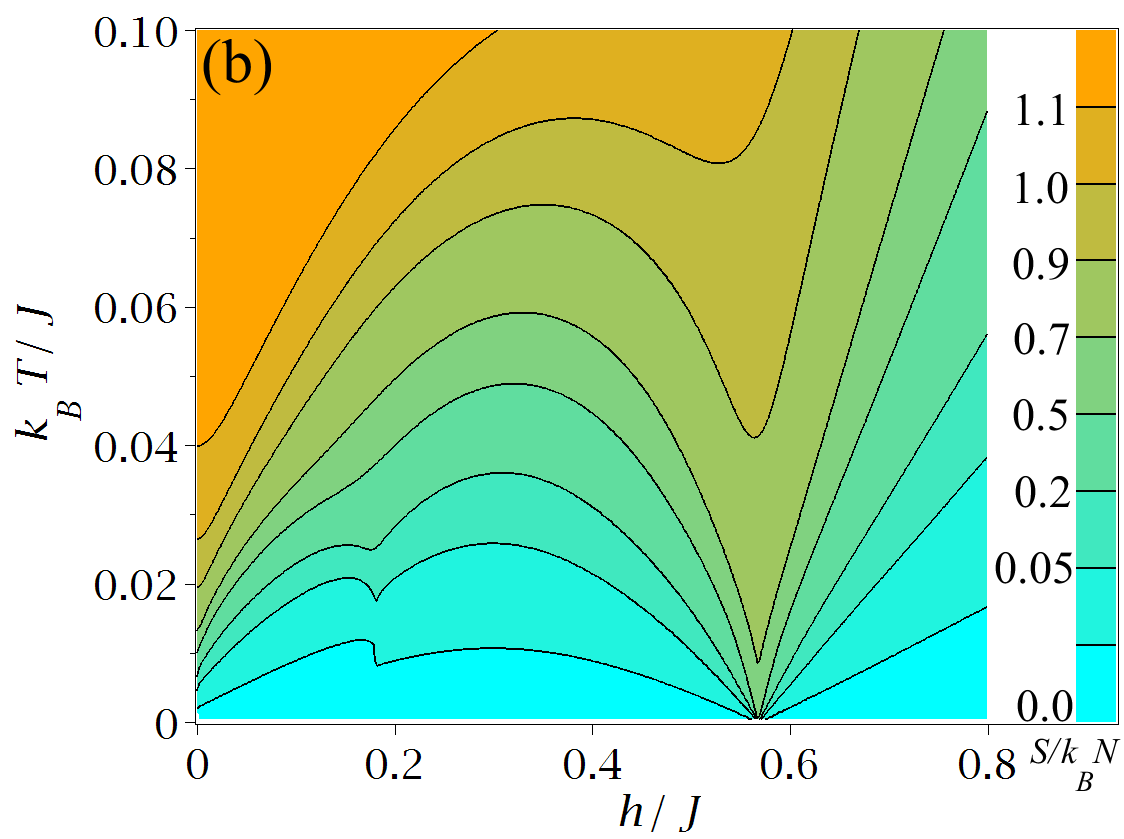}

\includegraphics[scale=0.1]{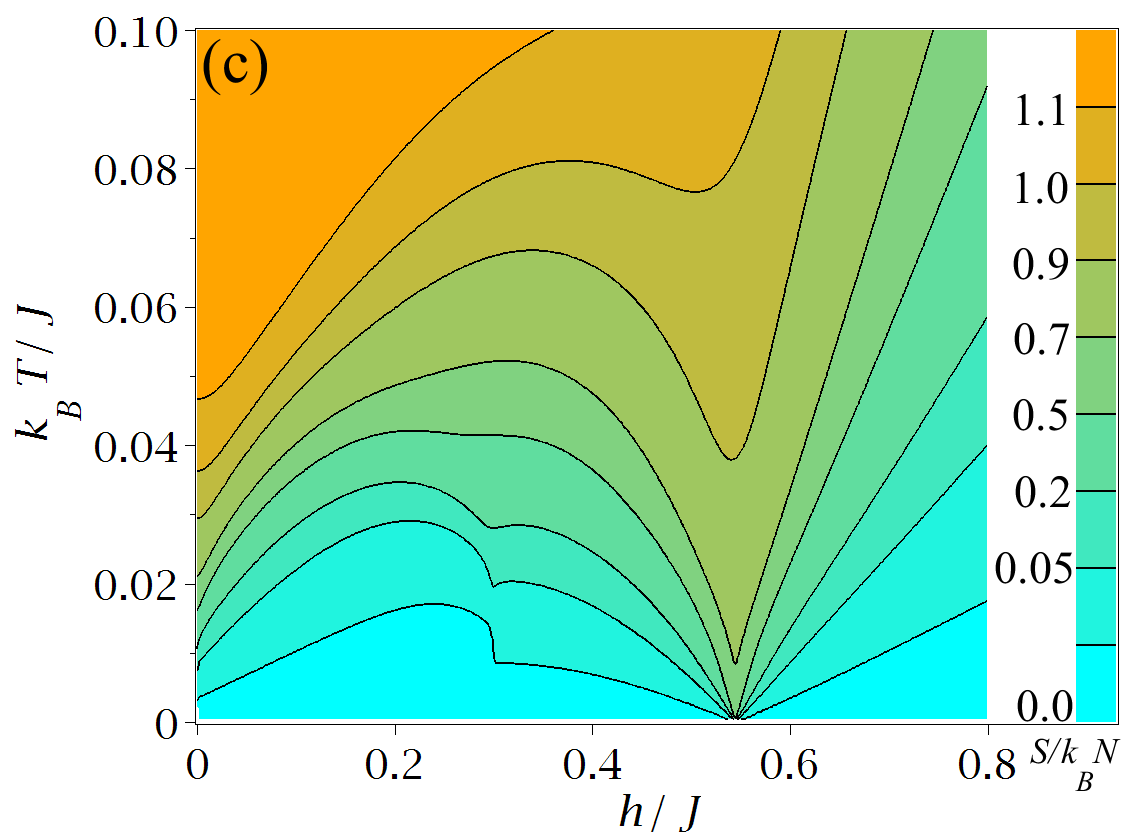}\includegraphics[scale=0.1]{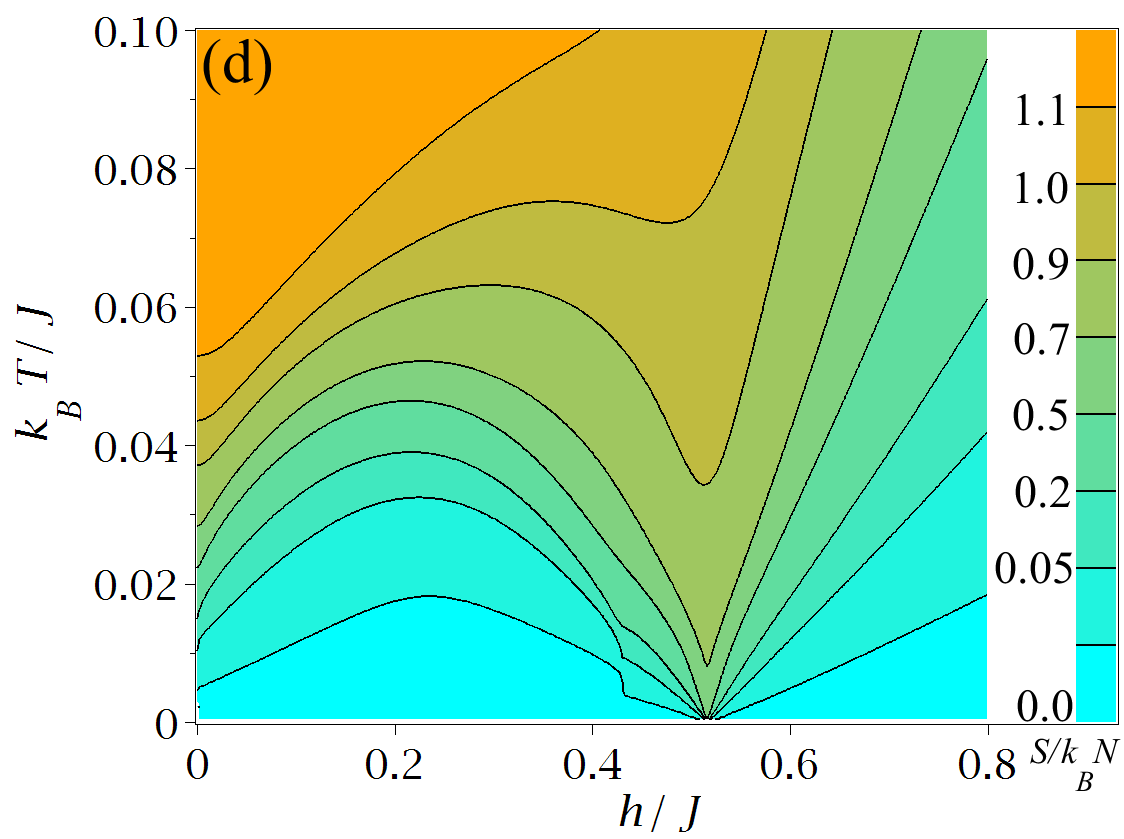}\caption{\label{fig:isentropy} (Color Online) Isentropic curve in the plane
temperature $k_{B}T/J$ versus magnetic field $h/J$, assuming $J_{0}/J$=-0.3. and $J_{z}/J=0.3$. (a) For fixed $\gamma=0.4$. (b) For fixed $\gamma=0.45$.
(c) For fixed $\gamma=0.5$. (d) For fixed $\gamma=0.55$. Each curves correspond to isentropy curve, from bottom to top are given for $\mathcal{S}=\{0.0001,0.05,0.2,0.5,0.7,0.9,1.0,1.1\}$.}
\end{figure}

In figure \ref{fig:isentropy} a typical isentropic change of temperature $k_{B}T/J$ is displayed varying the external magnetic field $h/J$, for two different magnetization scenarios discussed previously in fig.\ref{fig:Mag-XYZ}. In figure \ref{fig:isentropy}(a), we display the isentropic changes of temperature $k_{B}T/J$ as a function of the external magnetic field $h/J$, assuming fixed value of XY-anisotropy $\gamma=0.4$, $J_{z}/J=0.3$ and $J_{0}/J=-0.3$, where we can show one enhanced region of MCE at $h/J\approx0.6$, for several fixed entropies from bottom to top $\mathcal{S}=\{0.0001,0.05,0.2,0.5,0.7,0.9,1.0,1.1\}$. Despite at $h/J\approx0.1$, there is a small improvement, due to phase transition from $FMF_{2}$ to $FAF$ region. A similar plot is depicted in figure \ref{fig:isentropy}(b) for fixed value $\gamma=0.45$, and same set of parameters and fixed entropies given in figure \ref{fig:isentropy}(a), where is observed a similar enhanced MCE appearing at $h/J\thickapprox0.6$. Similarly, in Figures \ref{fig:isentropy} (c-d) also exhibits a similar behavior of the isentropic curve, considering the same set of parameters and entropies given as in figure \ref{fig:isentropy}(a), but for $\gamma=0.5$ and $\gamma=0.55$ respectively.

\begin{figure}
\includegraphics[scale=0.17]{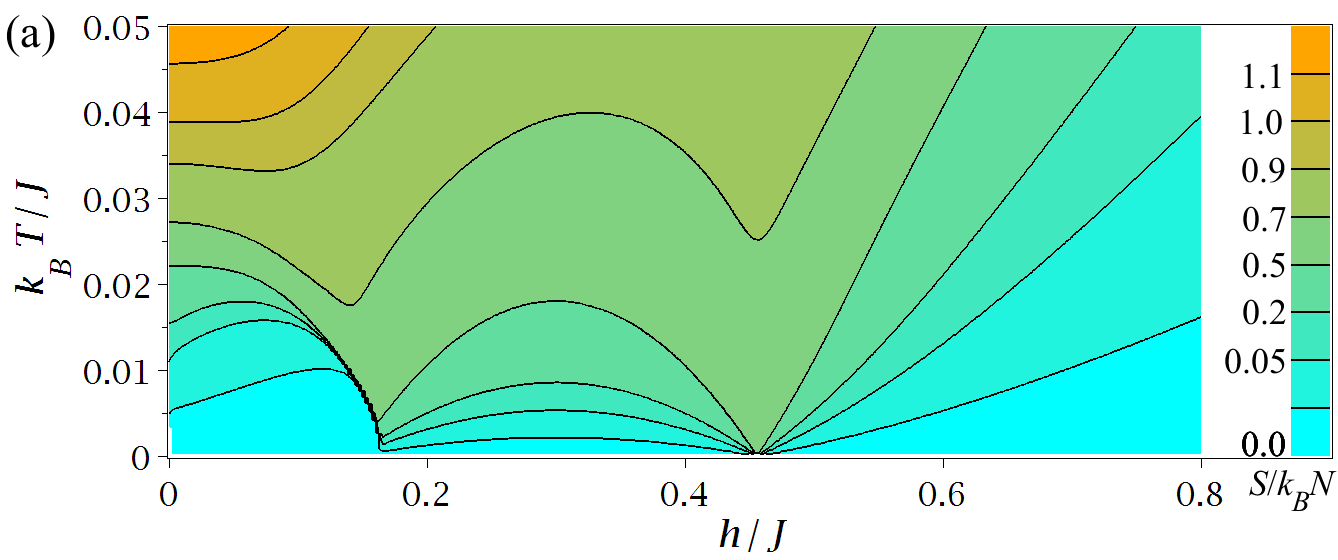}

\includegraphics[scale=0.17]{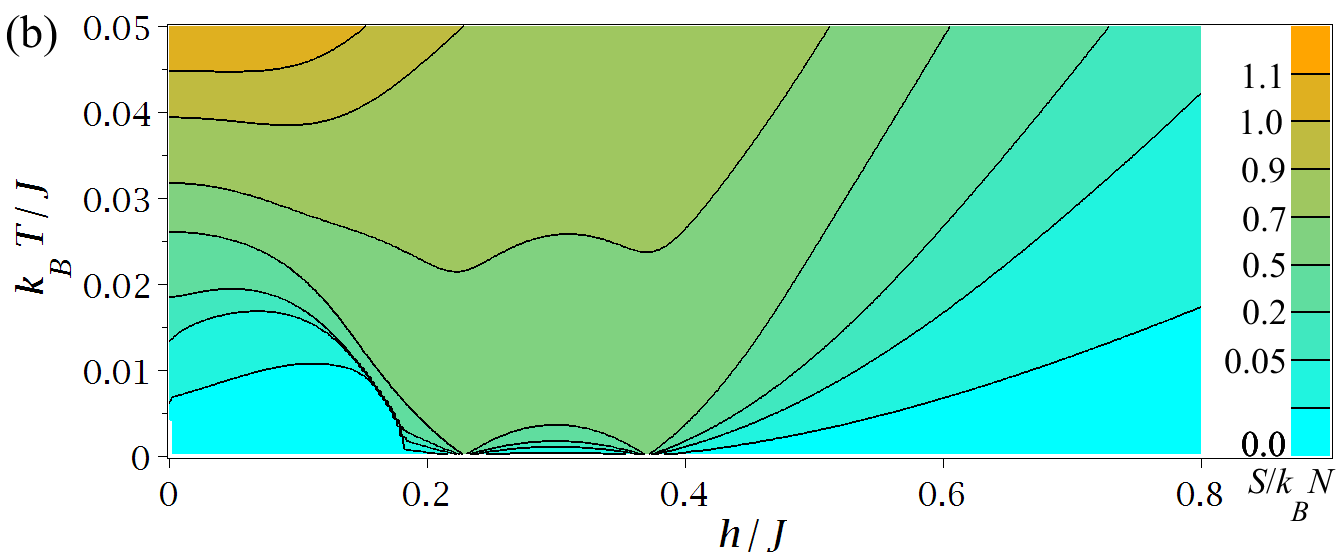}

\includegraphics[scale=0.17]{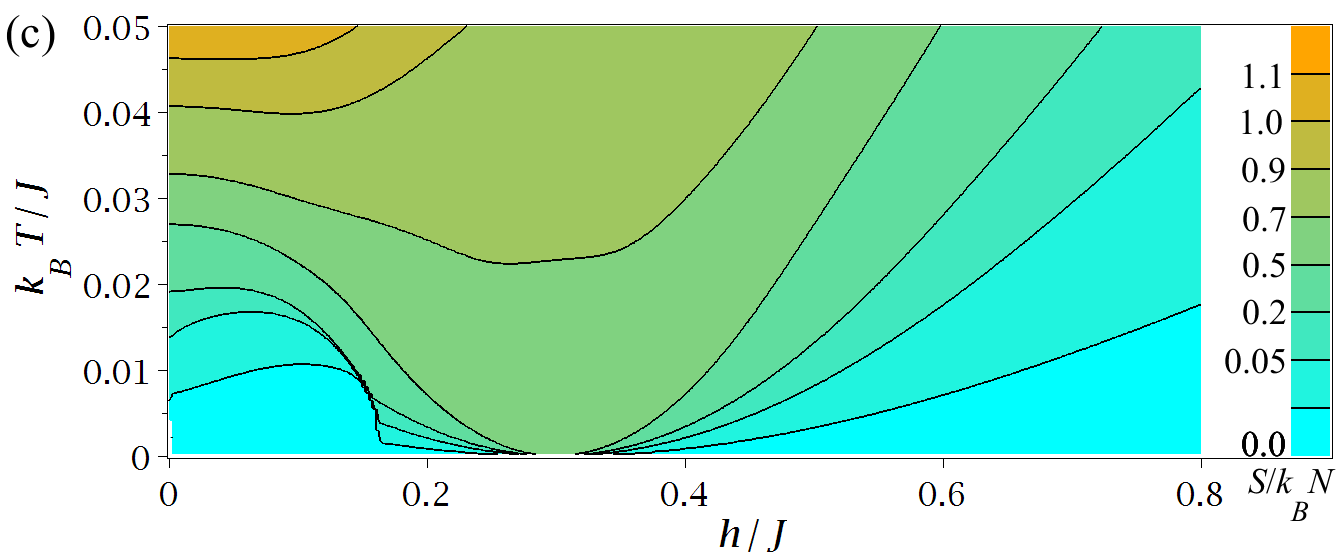}

\caption{\label{fig:isentrop2} (Color Online)Isentropic curve in the plane temperature $k_{B}T/J$ versus magnetic field $h/J$, for $J_{0}/J=-0.3$
and $J_{z}/J=0$. (a) $\gamma=0.95$, (b) $\gamma=0.99$, and (c) $\gamma=1.0$. Each curves correspond to isentropy curve, from bottom to top are given for $\mathcal{S}=\{0.0001,0.05,0.2,0.5,0.7,0.9,1.0,1.1\}$}
\end{figure}

Surely, from Fig. \ref{fig:isentrop2} one can observe the most relevant MCE, which can be observed just when the entropy is sufficiently close to $\mathcal{S}=k_{B}\ln(2)$, under which the temperature vanishes as soon as $\gamma\rightarrow1$, both phase transitions become closer to each other at $h/J\approx0.3$. In fig. \ref{fig:isentrop2}a is depicted for $\gamma=0.95$ and clearly we can observe two phase transitions points at $h/J\approx0.18$ and $h/J\approx0.48$ where the MCE could become enhanced. In fig. \ref{fig:isentrop2}b is illustrated for $\gamma=0.99$, the phase transitions occur at $h/J\approx0.24$ and $h/J\approx0.36$, here we can observe between both phase transitions there is a strong change in entropy, what means the MCE becomes more efficient than out of this interval. In fig.\ref{fig:isentrop2}c is shown for $\gamma=1.0$  and both phase transitions merged in a just one point $h/J=0.3$. Therefore, this phase transition leads to an enhanced MCE around $h/J\approx0.3$.

\begin{figure}
\includegraphics[scale=0.4]{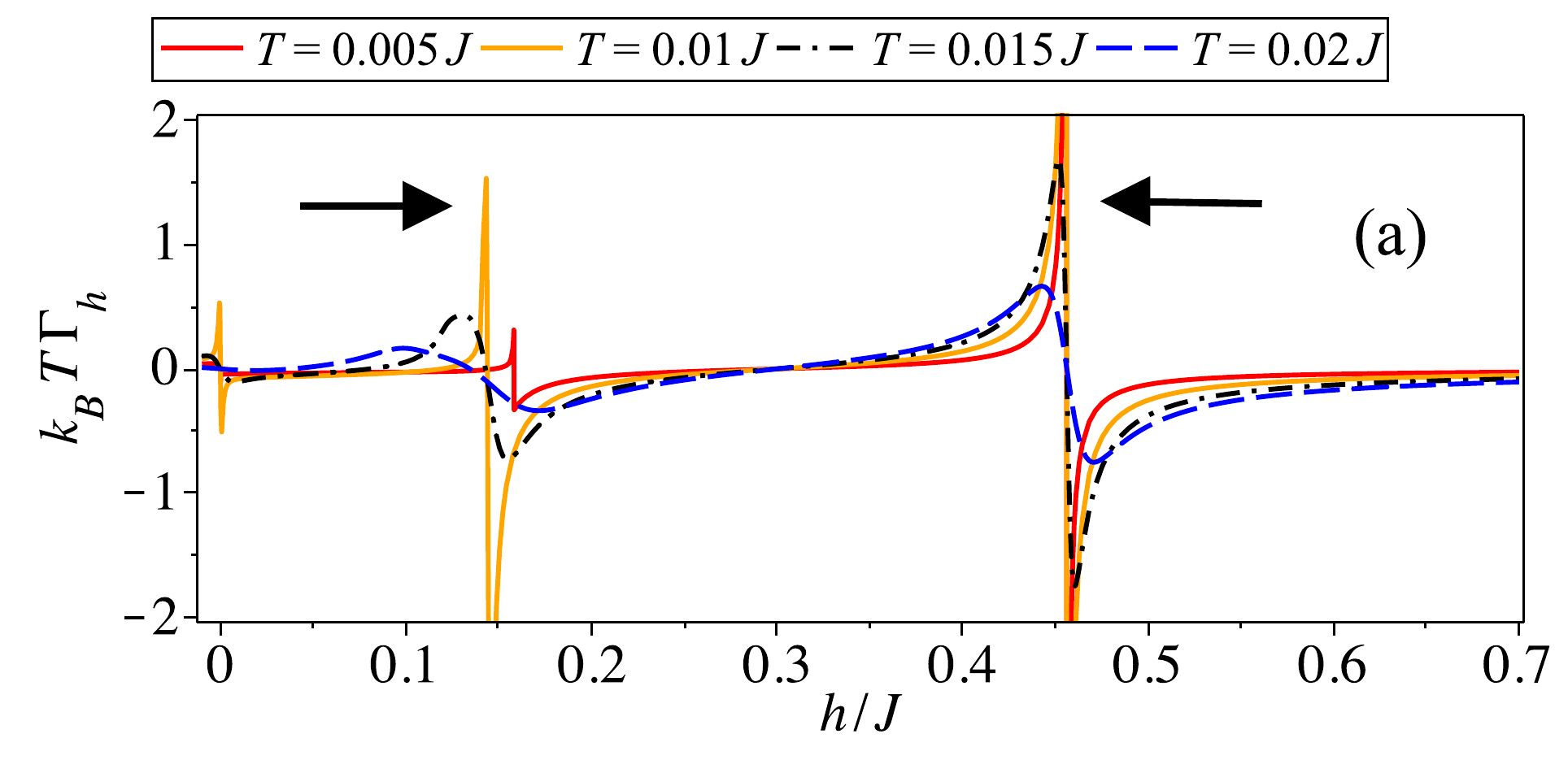}

\includegraphics[scale=0.4]{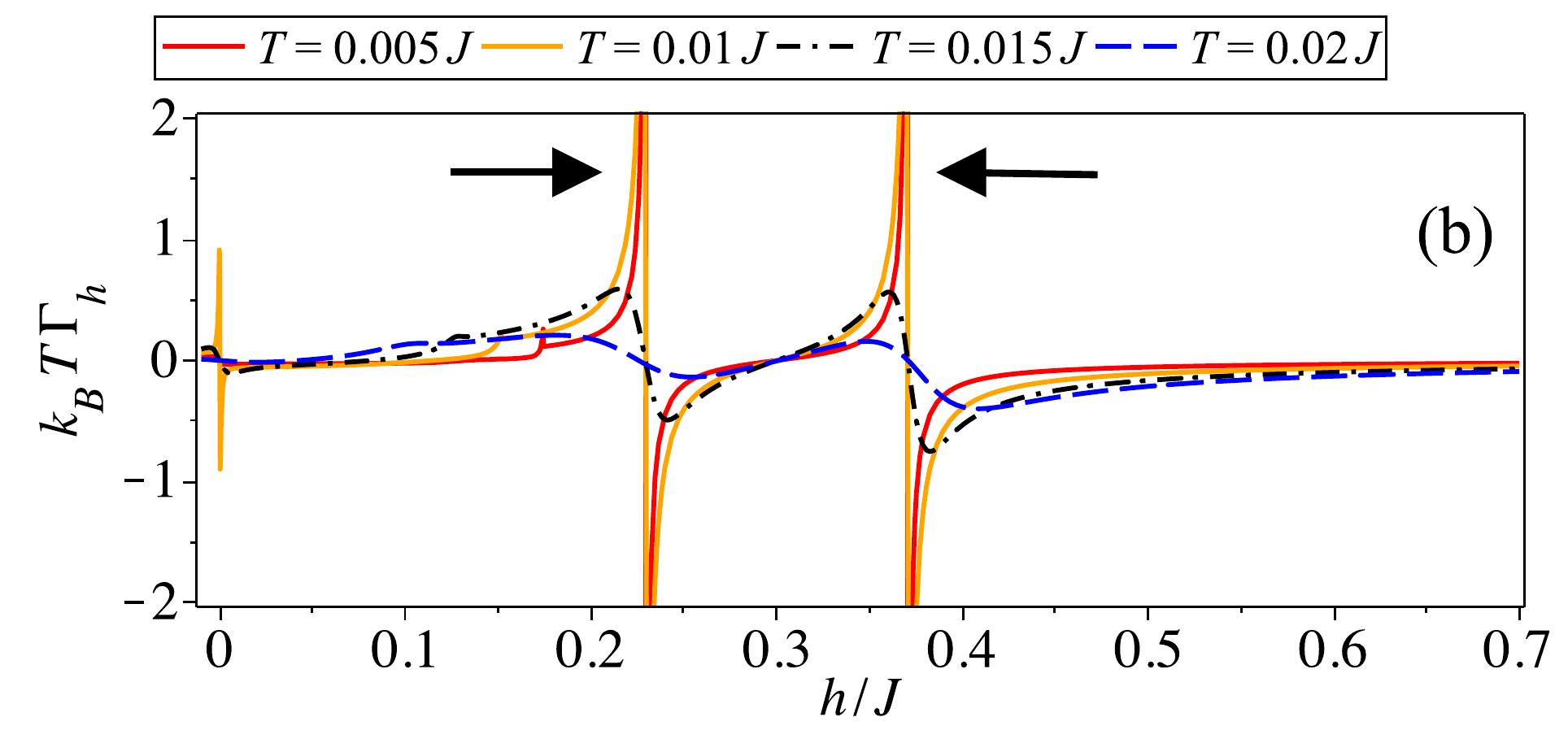}

\includegraphics[scale=0.4]{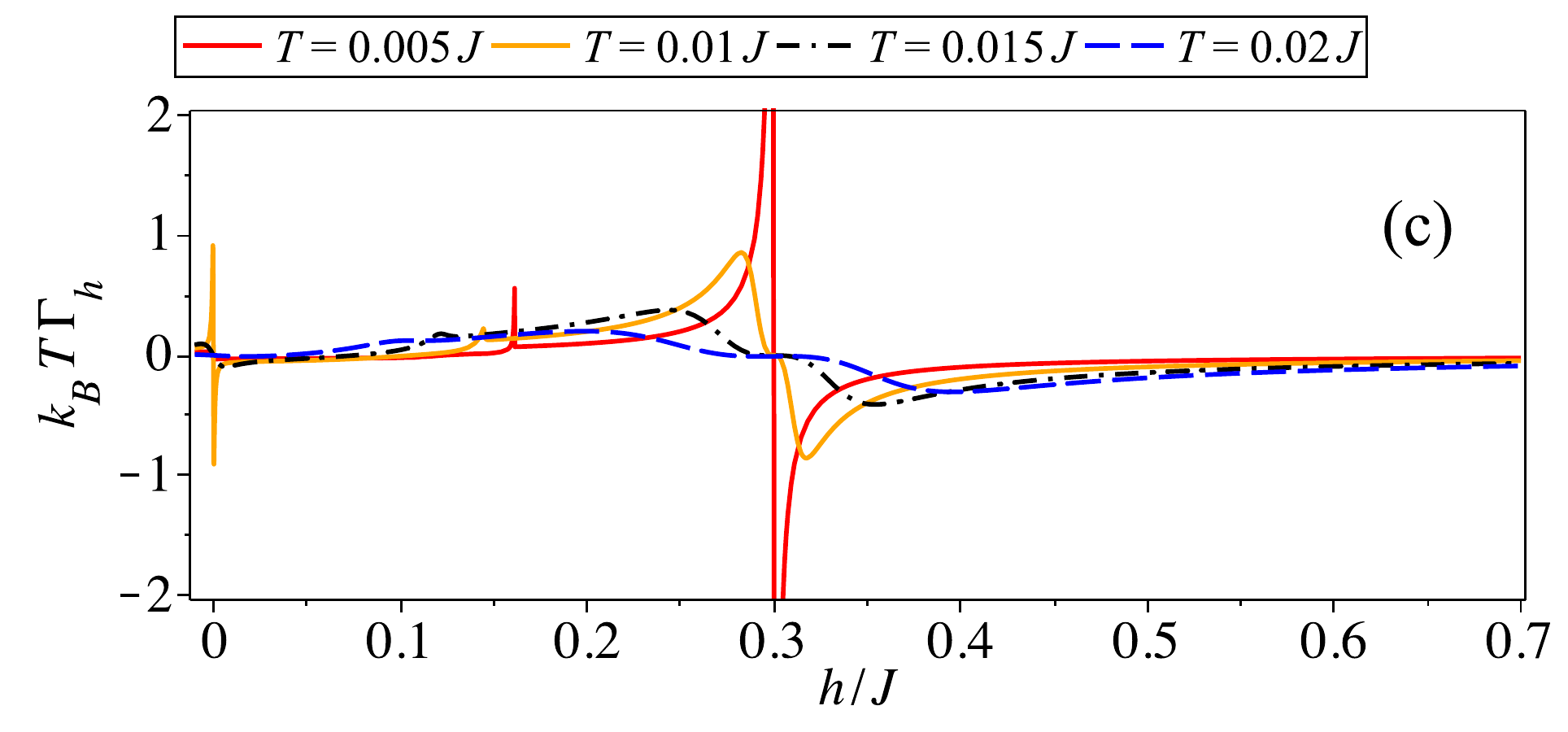}

\caption{\label{fig:The-Gr=00003D0000FC} (Color Online) The Grüneisen parameter $k_{B}T\Gamma_{h}$ as function of magnetic field $h/J$, for $J_{0}/J=-0.3$, $J_{z}/J=0$ and for different values of the temperature $T/J$. (a) $\gamma=0.95$, (b) $\gamma=0.99$, and (c) $\gamma=1.0$.}
\end{figure}

In the magnetocaloric effect ($\left(\frac{\text{\ensuremath{\partial}}T}{\text{\ensuremath{\partial}}h}\right)_{\mathcal{S}}=\text{\textminus}\frac{(\text{\ensuremath{\partial}}\mathcal{S}/\text{\ensuremath{\partial}}h)_{T}}{(\text{\ensuremath{\partial}}\mathcal{S}/\text{\ensuremath{\partial}}T)_{h}}$)
and the related quantity called the Grüneisen parameter $\Gamma_{h}$ has been pointed out as a valuable tool for detecting and classifying quantum critical points\cite{garst,Zhu}. Therefore, the Grüneisen parameter for magnetic systems under adiabatic conditions can be written as
\begin{equation}
\Gamma_{h}=\frac{1}{T}\left(\frac{\text{\ensuremath{\partial}}T}{\text{\ensuremath{\partial}}h}\right)_{\mathcal{S}}=-\frac{1}{C_{h}}\left(\frac{\partial M}{\partial T}\right)_{h},
\end{equation}
where $C_{h}$ is the specific heat at a constant magnetic field and $(\partial M/\partial T)_{h}$ is the temperature variation of the magnetization $M$. The Grüneisen parameter $\Gamma_{h}$ has a characteristic sign change close to the quantum critical point, which is due to the accumulation of entropy at the critical point\cite{garst}.

In figure \ref{fig:The-Gr=00003D0000FC}(a) we plot $k_{B}T\Gamma_{h}$ as a function of magnetic field $h/J$ for $J_{0}/J=-0.3$, in the low temperature region $0.005\leqslant k_{B}T/J\leqslant0.02$. There is a large Grüneisen parameter $k_{B}T\Gamma_{h}$ close to the phase transitions, the phase transition also corresponds to a frustrated region which is in agreement with the phase transition of figure 2 and 3 of reference \cite{torrico}. For XY-anisotropy $\gamma=0.99$ as shown in figure \ref{fig:The-Gr=00003D0000FC}(b), these peaks becoming closer to each other, keeping mainly the same kind of curves
compared to figure \ref{fig:The-Gr=00003D0000FC}(a). Nevertheless, it is valuable to observe that, for $T\lesssim0.01$, the curve shows a wider region where the MCE becomes efficient. Whereas, in figure \ref{fig:The-Gr=00003D0000FC}(c) we plot when $\gamma=1$, and these two peaks are merged in just a single peak at $h/J\approx0.3$. Surely, we could plot for other values of gamma such as considered in Fig.\ref{fig:isentropy}, but in those cases, we will retrieve the well known standard Grüneisen parameter's pick\cite{Trippe,galisova-2014,gali}.

\section{Conclusions}

In summary, we have presented a detailed study of the spin-1/2 Ising-XYZ chain on a diamond structure, which is an exactly solvable model by taking the decoration transformation and transfer matrix approach. In this work was discussed the phase diagram of ground state energy, displaying frustrated regions, as well as non-plateau magnetization at zero temperature. Due to XY-anisotropy parameter ($\gamma$) in the Hamiltonian, we found a remarkable interesting behavior for the magnetization, such as the presence of non-plateau magnetization, the XY-anisotropy influences in the arising of this phenomenon. Furthermore, due the existence of the non-plateau magnetization at zero temperature, we focus our discussion on the magnetocaloric effect (MCE) for the
present model, illustrating isentropic curves as well as the Grüneisen parameters. Consequently, when $\gamma\rightarrow1$ the two phase transitions just merged in one phase transition at $h/J=0.3$, thus, we observe that the MCE becomes more efficient in a wider interval of the magnetic field than in simple standard phase transition.

\section*{Acknowledgment}

O. Rojas, M. Rojas and S. M. de Souza thank Brazilian agency CNPq, FAPEMIG and CAPES for partial financial support. J. Torrico thanks CAPES for fully financial support.

\end{document}